%% file: main.tex
\documentclass[twocolumn]{aastex63}
\usepackage{verbatim}

\usepackage{booktabs} 
\usepackage{xparse}
\usepackage{xcolor}
\usepackage{amsmath}
\usepackage{filecontents,catchfile}

\NewDocumentCommand{\codeword}{v}{%
\texttt{\textcolor{black}{#1}}%
}

\accepted{in ApJ July 28, 2023}

\usepackage{lineno}

\begin{document}

\title{Tracing the Total Stellar Mass and Star Formation of High-Redshift Protoclusters}

\correspondingauthor{Roxana Popescu}
\email{rpopescu@umass.edu}

\author[0000-0001-8245-7669]{Roxana Popescu}
\affiliation{Department of Astronomy, University of Massachusetts, 710 North Pleasant Street, Amherst MA 01003, USA}

\author[0000-0001-8592-2706]{Alexandra Pope}
\affiliation{Department of Astronomy, University of Massachusetts, 710 North Pleasant Street, Amherst MA 01003, USA}

\author[0000-0003-3004-9596]{Kyoung-Soo Lee}
\affiliation{Department of Physics and Astronomy, Purdue University, 525 Northwestern Avenue, West Lafayette, 
IN 47907, USA}

\author[0000-0002-8909-8782]{Stacey Alberts}
\affiliation{Steward Observatory, University of Arizona, 933 N. Cherry Street, Tucson AZ, 85721 USA}

\author[0000-0001-6320-261X]{Yi-Kuan Chiang}
\affiliation{Center for Cosmology and AstroParticle Physcis (CCAPP), The Ohio State University, Columbus, OH 43210, USA}
\affiliation{Institute of Astronomy and Astrophysics, Academia Sinica (ASIAA), Taipei 10617, Taiwan}

\author[0000-0002-2127-6060]{Sowon Lee}
\affiliation{Department of Physics and Astronomy, Purdue University, 525 Northwestern Avenue, West Lafayette, 
IN 47907, USA}

\author[0000-0002-4208-798X]{Mark Brodwin}
\affiliation{Department of Physics and Astronomy, University of Missouri, 5110 Rockhill Road, Kansas City, MO 64110, USA}

\author[0000-0002-6149-8178]{Jed McKinney}
\affiliation{Department of Astronomy, University of Massachusetts, 710 North Pleasant Street, Amherst MA 01003, USA}
\affiliation{Department of Astronomy, The University of Texas at Austin, 2515
Speedway Blvd Stop C1400, Austin, TX 78712, USA}

\author[0000-0002-9176-7252]{Vandana Ramakrishnan}
\affiliation{Department of Physics and Astronomy, Purdue University, 525 Northwestern Avenue, West Lafayette, 
IN 47907, USA}

\begin{abstract}

As the progenitors of present-day galaxy clusters, protoclusters are excellent laboratories to study galaxy evolution. Since existing observations of protoclusters are limited to the detected constituent galaxies at UV and/or infrared wavelengths, the details of how typical galaxies grow in these young, pre-virialized structures remain uncertain. We measure the total stellar mass and star formation within protoclusters, including the contribution from faint undetected members by performing a stacking analysis of 211 $z=2-4$ protoclusters selected as Planck cold sources. We stack WISE and Herschel/SPIRE images to measure the angular size and the spectral energy distribution of the integrated light from the protoclusters. The fluxes of protoclusters selected as Planck cold sources can be contaminated by line of sight interlopers. Using the WebSky simulation, we estimate that a single protocluster contributes $33\pm15$\% of the flux of a Planck cold source on average. After this correction, we obtain a total star formation rate of $7.3\pm3.2 \times 10^3\ M_{\odot} {\rm yr}^{-1}$ and a total stellar mass of $4.9\pm 2.2\times 10^{12}\ M_{\odot}$. Our results indicate that protoclusters have, on average, 2x more star formation and 4x more stellar mass than the total contribution from individually-detected galaxies in spectroscopically-confirmed protoclusters. This suggests that much of the total flux within $z=2-4$ protoclusters comes from galaxies with luminosities lower than the detection limit of SPIRE ($L_{IR} < 3 \times 10^{12} L_{\odot}$). Lastly, we find that protoclusters subtend a half-light radius of 2.8\arcmin\ (4.2--5.8 cMpc) which is consistent with simulations.

\end{abstract}

\section{Introduction} 
\label{sec:intro}

Protoclusters are overdensities of galaxies at high redshift that are expected to collapse into low redshift galaxy clusters \citep{2016A&ARv..24...14O}. Unlike galaxy clusters, protoclusters are not yet compact and virialized. In the local universe, galaxies located within galaxy clusters have lower star formation rates and older stellar populations compared to galaxies in the field \citep{1980ApJ...236..351D}. However, observations and simulations of galaxies suggest that at higher redshifts cluster galaxies have higher star formation rates for a fixed stellar mass \citep{2016ApJ...824...36C, 2017ApJ...844L..23C}. At redshifts of $z>1.2$, the star-formation rates of individual galaxies within clusters exceed those of field galaxies \citep{2014MNRAS.437..437A}. 

It is important to study protoclusters to understand the history of the formation of galaxy clusters and the processes that govern the growth and quenching of galaxies that reside in dense environments. Studies of protoclusters can also indicate how structure evolves in the universe and indicate the impact of environment on galaxy evolution. Infrared observations, in particular, provide a powerful constraint on stellar content growth via star formation in protoclusters \citep[see, e.g.,][and references therein]{2022Univ....8..554A}. At $z>2$, continuum emission from old stars is redshifted to the near-infrared. Dust obscured star-formation is observed in the mid- to far-infrared as emission from stars is absorbed by and re-radiated from the dust. State-of-the-art hydrodynamic simulations are unable to reproduce the combined star-formation rates of highly dust-obscured starburst galaxies observed in protoclusters \citep{lim21,2022arXiv220801053R}. Sensitive infrared observations can place firm observational constraints on how galaxies grow in protocluster environments and inform theoretical models.

Since protoclusters are extended structures that have not yet collapsed, the hot intracluster gas might not exist. As a result, conventional methods of detecting galaxy clusters such as the Sunyaev-Zeldovich effect and X-ray emission are not as effective. In recent years, high-redshift protoclusters have been discovered as overdensities of Ly$\alpha$ emitters \citep[LAEs: e.g.,][]{2000ApJ...532..170S,2000A&A...358L...1K,2004AJ....128.2073H,2014ApJ...796..126L,2016ApJ...823...11D,2019ApJ...879...28H,2019ApJ...879....9S}, H$\mathrm{\alpha}$ emitters \citep{2004A&A...428..793K,2011PASJ...63S.415T,2012ApJ...757...15H,2013MNRAS.428.1551K}, Lyman Break Galaxies \citep[LBGs: e.g.,][]{1998ApJ...492..428S,2012ApJ...750..137T,2019ApJ...871...83S},  dusty star-forming galaxies  \citep[DSFGs: e.g.,][]{2009ApJ...691..560C,2018ApJ...856...72O,2015ApJ...808L..33C} and through infrared color selection \citep[e.g.,][]{2012ApJ...749..169G}. Recently, observatories designed to map the CMB such as Planck and the South Pole Telescope (SPT) have identified bright millimeter/submillimeter sources, which, when followed up with observations at higher resolution, resolve into overdensities of dusty galaxies that are candidate protoclusters \citep{2015A&A...582A..30P,2021MNRAS.508.3754W,2018Natur.556..469M}.  While these selection techniques have been proven effective in identifying protoclusters, they target specific types of galaxies and thus do not provide a complete picture of protocluster formation \citep{2013ApJ...778..170K,2019ApJ...871...83S}. Additionally, the small number of deep, multi-wavelength follow-up observations of protoclusters do not present a uniform census of the low-luminosity members that may dominate in number and in the star-formation/stellar mass budget.

In principle, confirming a protocluster structure requires spectroscopic measurements of its member galaxies. However, even in the era of deep, wide-field imaging surveys by facilities such as Euclid, Vera C. Rubin Observatory, and the Nancy Grace Roman Space Telescope which will identify a large number of protocluster candidates, it will be expensive to spectroscopically follow up all of these candidates. A study of a statistical sample of protoclusters that accounts for all of the emission in protocluster members and not just the massive and bright galaxies is needed to quantify when and where star formation takes place within protoclusters and to provide a useful benchmark to interpret further observations. A statistical study would also be useful for comparing the average properties of protoclusters selected by different methods. Comparing statistical samples of protocluster candidates identified by overdensities in the far-IR to optically or near-IR selected overdensities would demonstrate differences in the total star formation within these populations.

In this study we use a WISE and Herschel\footnote{Herschel is an ESA space observatory with science instruments provided by European-led Principal Investigator consortia and with important participation from NASA.} SPIRE stacking analysis to measure the total light from Planck-identified protocluster candidates \citep{2015A&A...582A..30P}  in order to quantify the average stellar mass and the dust-obscured star formation rate. We compare our results to a similar stacking analysis done by \cite{2019ApJ...887..214K} on protoclusters detected by overdensities of Lyman break galaxies with the Hyper Suprime-Cam Subaru Strategic Program \citep{2018PASJ...70S..12T}. We also compare to clusters at $z\sim1-2$ \citep{2021MNRAS.501.1970A} stacked using the same total light stacking method.

This paper is organized as follows: Section \ref{sec:data} describes the protocluster sample and the datasets used in this work. In Section \ref{sec:method} we discuss our total light stacking technique. In Section \ref{subsec:los} we calculate a correction factor for line of sight interlopers within the sample.  In Section \ref{sec:res}, we apply the stacking technique to the Planck protocluster candidates, analyzing the average radial profiles and total photometry. We build SEDs from measurements of the light from the stacked protoclusters. Section \ref{sec:disc} discusses our results and Section \ref{sec:conclusion} presents our conclusions. Throughout this work, we adopt the standard cosmology ($\Omega_{M}$, $\Omega_{\Lambda}$, \textit{h}) = (0.3, 0.7, 0.70)  and a \cite{2003PASP..115..763C} initial mass function (IMF).

\section{Data}
\label{sec:data}

\subsection{Protocluster Sample}

Our sample consists of protocluster candidates from \cite{2015A&A...582A..30P} that were selected as ``cold" sources in the cosmic infrared background from Planck High Frequency Instrument (HFI) maps or selected from the Planck Catalogue of Compact Sources (PCCS: \citealt{2014A&A...571A..28P}). A ``cold" source has a redder color, peaking between 353 and 857 GHz, which corresponds to a cold dust temperature or a high redshift. We hereafter refer to this sample as the PC15 protocluster candidates.

According to \cite{2016A&A...596A.100P} the sources in the Planck High-z (PHz) catalog, which were selected similarly to the PC15 protocluster candidates, have far-infrared luminosities ranging from $ 1-10 \times 10^{14}L_{\odot}$ and a surface density of 0.2~deg$^{-2}$. A cold point source identified with Planck, which has an extent smaller than the 4\arcmin.5 Planck beam, is likely an overdensity of unresolved dusty sources or a strongly lensed galaxy \citep{2018MNRAS.476.3336G,2005MNRAS.358..869N,2017MNRAS.470.2253N,2022A&A...664A.155G}. As described in \cite{2015A&A...582A..30P}, 228 Planck sources identified were followed up with Herschel SPIRE, which has a smaller beamsize than Planck and can resolve the Planck sources into overdensities or single lensed galaxies. Of the 228 sources, 12 are strongly lensed galaxies confirmed by spectroscopic follow-up, 4 are dominated by Galactic cirrus and the remaining 212 are robust protocluster candidates.  Stacked observations of these protocluster candidates show significant and extended emission that extends larger than the Planck beam \citep{2015A&A...582A..30P}. We assume a redshift of $z=2$ for the PC15 protoclusters, since this is consistent with the colors of individual SPIRE sources, assuming a dust temperature of 35 K \citep[Figure 13,][]{2015A&A...582A..30P}.

51 protocluster candidates from both the PC15 protocluster candidates as well as other protocluster candidates from the PHz catalog were followed up with observations by SCUBA-2 on the JCMT \citep{2017MNRAS.468.4006M}. They find that within the combined 1.20 $\mathrm{deg^2}$ of these 51 fields there is an enhanced source density of $S_{850} > 8$~mJy sources relative to the field. \cite{2020MNRAS.494.5985C} individually analyze 46 of these 51 fields and classify 25 as significantly overdense, 11 as slightly overdense, and 10 as not overdense. \cite{2018A&A...620A.198M} follow up 83 of the PC15 and PHz protocluster candidates with Spitzer IRAC and determine that around 46\% of the fields are $3\sigma$ overdense compared to the field. Three individual Planck protocluster candidates were spectroscopically observed to identify member galaxies \citep{2016A&A...585A..54F,2019A&A...625A..96K,2021A&A...654A.121P}. One of these protocluster candidates is observed to be an overdensity of H$\alpha$-emitter galaxies \citep{2021MNRAS.503L...1K}. \cite{2022A&A...662A..85P} follows up 18 PHz fields with IRAM-30m/EMIR to detect CO emission from individual galaxies. Out of 8 IRAM/EMIR fields with successfully measured redshifts, half contain 2--3 galaxies each with similar redshifts. In summary, follow-up observations of Planck protocluster candidates so far indicate that at least half correspond to overdense fields.

\subsection{Herschel Imaging}

From the Herschel Science Archive, we download the level 2  data products of 211 PC15 protocluster candidates observed with Herschel/SPIRE under the program IDs,
{\tt OT1\_lmontier\_1}, {\tt OT2\_hdole\_1}, and {\tt DDT\_mustdo\_5} (H. Dole, private communication). The data include all three SPIRE bands centered on 250, 350, and 500 $\mu$m. The dimensions of individual SPIRE images for this sample range from 24\arcmin\ to 44\arcmin\ for each band. The FWHM of the SPIRE beam is 18.1\arcsec, 24.9\arcsec, 36.6\arcsec\ for SPIRE 250, 350 and 500 $\mu$m bands respectively. The pixel scale of the images is 6\arcsec, 10\arcsec, and 14\arcsec\ for SPIRE 250, 350 and 500 $\mu$m bands respectively \citep{2010A&A...518L...3G}. The SPIRE confusion limit  is $1\sigma \sim$ 5.8, 6.3 and 6.8~mJy/beam at 250, 350 and 500~$\mu$m,  respectively \citep{2010A&A...518L...5N}.

\subsection{WISE Imaging}

The WISE mission provides all-sky coverage with four bands that are centered on 3.4 $\mu$m ($W1$), 4.6 $\mu$m ($W2$), 12 $\mu$m ($W3$), and 22 $\mu$m ($W4$) taken by several cryogenic and post-cryo surveys  \citep{2010AJ....140.1868W,2011ApJ...731...53M}. The FWHM of the WISE beam is 6.1\arcsec, 6.4\arcsec, 6.5\arcsec, 12.0\arcsec, for the WISE $W1$, $W2$, $W3$, $W4$ bands, respectively. In this work, we  use a set of coadds known as the unWISE dataset \citep{2014AJ....147..108L,2017AJ....153...38M}, specifically the unWISE `NEOWISE-R6' images \citep{2017AJ....154..161M}.  Though the unWISE coadds use the same set of observations as the official ALLWISE release \citep{cutri13}, they preserve the instrument PSF and the native pixel scale of 2.75\arcsec~$\mathrm{pix^{-1}}$ and do not apply additional smoothing.

We use the masked coadds in which outlier pixels are rejected when combining the images. We download tiles from the unWISE website that contain the center of each protocluster candidate. Each WISE tile is 2048 pixels on a side, covering a $1.56^\circ \times 1.56^\circ$ area.

We determine which tile is best to download for each protocluster by identifying the tile with the smallest separation between the tile center and the protocluster center. We  inspect the unWISE tiles containing the protocluster candidates. Some lie near the intersection of two or four adjacent WISE tiles. In principle, all of these frames contain useful data. However, using multiple tiles for a given protocluster can introduce unwanted systematic effects. As each WISE tile is projected with its own field center as a tangent point, simply coadding two adjacent tiles would artificially decrease the sky noise. In addition, the unWISE data processing includes its background subtraction \citep{lang16a} and  these overlapping tiles often have different background levels, which can potentially dilute the low-level signal.

To minimize this effect, we reject the unWISE tiles if the center of a given cluster lies too close to the image boundaries.  We require that a protocluster center lies within 0.7$^\circ$ in both X and Y directions from the center of the tile, ensuring that it is at least 7.5\arcmin\ away from the image boundary. 
20 protoclusters do not meet this criterion and thus are removed from subsequent analysis. Additionally, our visual inspection reveals that  some frames contain a star or an image artifact that is too close to the protocluster. In a handful of frames, one bright star dominates a substantial fraction of the frame with its scattered light, raising the background level for a much wider surrounding area, which could mimic the protocluster signal. All in all,  59 protoclusters are removed from the WISE analysis leaving 152 protoclusters.  We do not expect the removal of any frames to bias the WISE image stacks since the properties of the removed protoclusters are not intrinsically different from the rest of the sample.

\section{Protocluster Stacking Method}
\label{sec:method}

\begin{figure*}[ht!]
\centering
\includegraphics[width=6in]{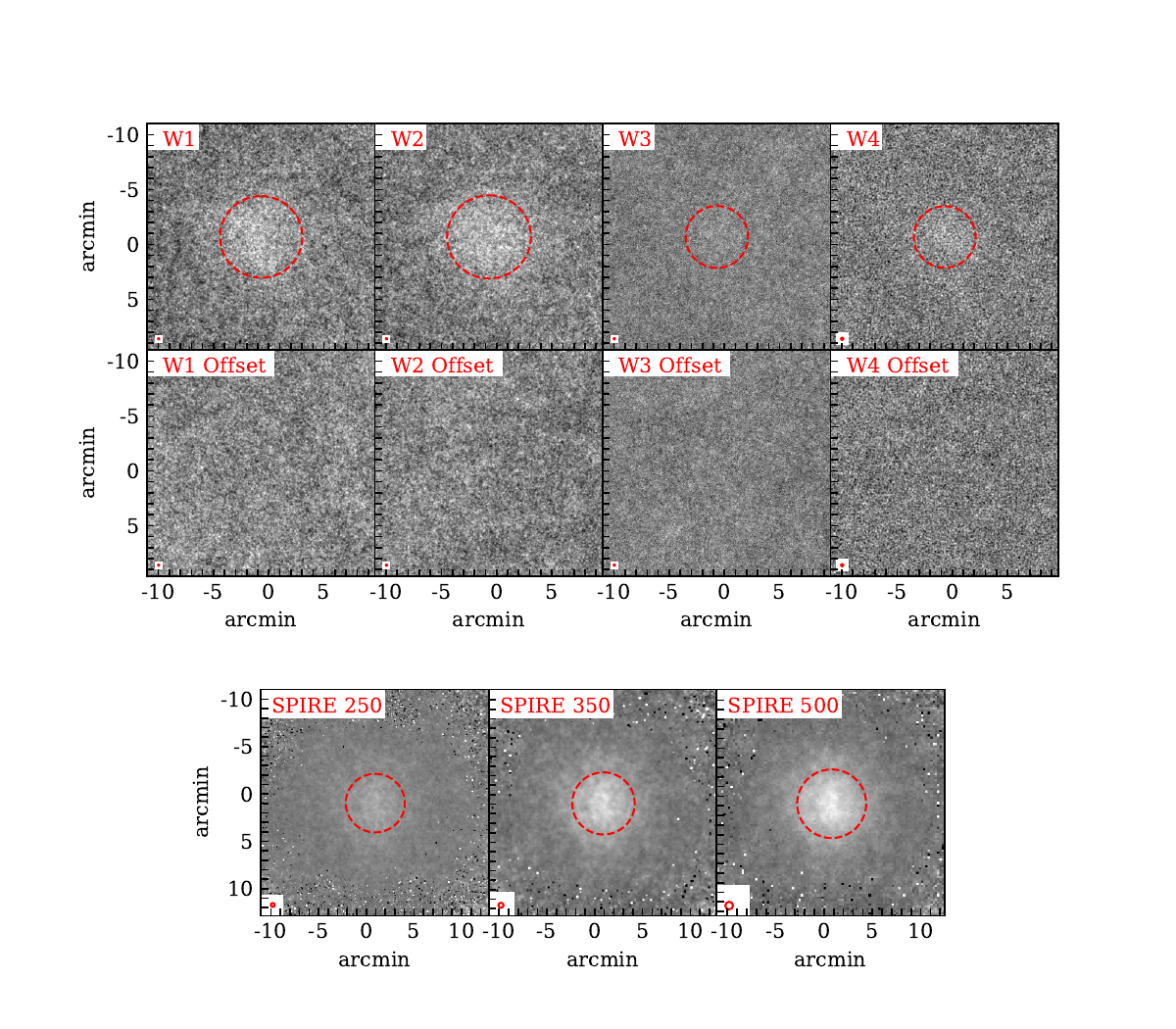}
\caption{Weighted mean stacks of the PC15 protoclusters showing extended emission from 3.4 $\mathrm{\mu m}$ - 500 $\mathrm{\mu m}$. The top panel shows the stacks and the offset stacks in the WISE bands. The offset stacks are created to verify that the extended emission does not appear when stacking random areas of the sky background. The bottom panel shows the stacks in the SPIRE bands. The size of the image point spread function FWHM is indicated by a red circle at the bottom left corner of each image. The measured half-light radii of the protoclusters are indicated by the dashed red circles.}
\label{stack_fig}
\end{figure*}

For our analysis we use the same total light stacking method developed in \cite{2021MNRAS.501.1970A} which captures detected cluster members as well as undetected lower luminosity cluster members and/or intra-cluster dust (ICD). In this paper, we stack WISE and SPIRE images of the PC15 protocluster candidates. We create cutouts around each of the PC15 protocluster candidates and stack the cutouts to measure the average signal from all constituents within a sample of protocluster candidates. In this section, we describe the processing of the WISE and Herschel data and our stacking analysis.
 
\subsection{WISE Image Processing}

\label{sec:wise_processing}

We first remove all individually detected sources from these images, as we expect most detected sources to be foreground galaxies and stars, since only very massive galaxies would be detected in WISE at $z=2$. The $5\sigma$ limiting magnitude of the unWISE $W1$, $W2$, $W3$, and $W4$ data is 19.6, 19.3, 16.8, and  14.7~AB, respectively \citep{2010AJ....140.1868W,cutri13,2017AJ....153...38M}. 
 Assuming the  Chabrier IMF and the formation redshift of $z_f=6$, a WISE source at $z=2$ detected in W1 with S/N=5 is expected to have a stellar mass of $\approx 4 \times 10^{11}\ M_{\odot}$, $3 \times 10^{11}\ M_{\odot}$, and $2.2 \times 10^{11}\ M_{\odot}$ for an exponentially decaying star formation history (SFH) with $\tau$ = 0.1, 1.0 Gyr, and a constant SFH, respectively. The use of the Salpeter IMF would increase the mass by 50--60\% while changing the formation redshift to $z_f=4$ would decrease the mass by up to 30\%. Given everything equal, a higher source redshift would also increase the mass of the detected galaxy.
 
Based on studies of DSFG overdense protoclusters located at $z=2-3$ \citep{2021A&A...654A.121P,2015ApJ...808L..33C,2015ApJ...799...38K}, we can expect that there are around 3 spectroscopically confirmed galaxies per protocluster field with WISE detections and stellar masses greater than $10^{11} M_{\odot}$. By masking out the WISE detections in our analysis, we might be losing the flux from at most 3 galaxies per protocluster field, which does not affect the stellar mass derived significantly (see Section \ref{sec:sed}).

 In order to isolate the much fainter signal from the protoclusters, we create pixel masks  as follows. First, we use the SExtractor software \citep[][SE, hereafter]{1996A&AS..117..393B} to detect sources in each WISE tile. We use a {\tt DETECT\_THRESH} of 3.0 and a {\tt DETECT\_MINAREA} of 3 pixels, requiring the isophotal signal-to-noise ratio to be 5 or higher.  While changing the detection setting can alter the overall background level in the final stacked image, our results are insensitive to such changes. The detection check image is generated as a segmentation map that tags all pixels belonging to an individual source.
 
 Second,  we estimate a constant background of each $1.5^\circ \times 1.5^\circ$ WISE tile and subtract it from the image. This step is different from the approach taken by \citet{2021MNRAS.501.1970A}, in which we used the SE-generated, spatially varying background map with {\tt BACK\_SIZE} of 10\arcmin.
 At $z=1.0$ (1.5), the virial radius of the galaxy clusters studied by \citet{2021MNRAS.501.1970A} is 1~Mpc corresponding to 2.0\arcmin\ (1.9\arcmin), i.e, much smaller than the scale of 10\arcmin\ at which the background is allowed to vary. In comparison, unvirialized protoclusters can have half-light radii up to 10~cMpc \citep[6.5\arcmin\ at $z=2$: e.g.,][]{2013ApJ...779..127C,muldrew15}. Although the fraction of pixels in a $1.5^\circ \times 1.5^\circ$ WISE tile containing a single protocluster is still expected to be small, allowing the background to vary within the tile could remove the low-level protocluster signal. To avoid this effect, we estimate the sigma-clipped mode value  and subtract it from the entire image.

 Finally, using the segmentation map we mask all pixels that belong to individually detected WISE sources and saturated pixels. The masked pixels are flagged as NaN. The mean (median) fraction of pixels flagged as NaN for the full sample is 31\% (26\%) with a standard deviation of 13\% for $W1$. As for $W4$, the number is 13 \% (7\%).

\subsection{WISE Image Stacking}\label{subsec:wise_stacking}

For each protocluster, we create a 901$\times$901 pixel image (41\arcmin\ on a side) centered on its position using the IDL routine {\tt hastrom.pro}. All pixels in the no-data region are replaced with NaN.
We then convert counts to physical units and store the resampled image of each protocluster in a three-dimensional datacube. We create a weighted mean stack of the datacube where we use the inverse variance of the pixel-wise rms as the weight of each protocluster image.

The stacked image inherently contains residual background which primarily originates from faint sources that are not masked prior to stacking. 
Additionally, the averaging of $>100$ images reveals the low-level residual sky that had not been properly accounted for at the initial background subtraction step. We estimate the constant background in the annular bins in the range of 9\arcmin--12\arcmin. At $z=2$ (3), the range corresponds to 14--18~cMpc (17--23~cMpc), and thus the range ensures that the background is estimated safely away from the expected extent of protoclusters.

It is worth stressing that  rigorous sky subtraction is key to obtaining a robust stacking result. It is particularly important for protocluster analyses not only  because the expected signal is diffuse but also because protoclusters are very extended in the sky. The use of an SE-generated spatially-varying background map for background subtraction instead of a constant background can considerably alter the signal. Setting the spatial filter size (set by the {\tt BACK\_SIZE} parameter in SE) to be  larger than the expected protocluster size mitigates the problem. Any filter smaller than $\approx 15$\arcmin\ has the potential to remove the low-level signal from the protocluster and to alter the light profile. 

Still, the resultant stacked images suggest that the background may have been oversubtracted. Visual inspection of the stacked image -- most prominent in $W1$ --  shows a slight increase in the background level at $>12$\arcmin\ from cluster center. This upturn does not depend on the manner in which we perform sky subtraction including the choice of {\tt BACK\_SIZE} values. Thus, we speculate that it is a feature already present in the unWISE tiles. 
As described in \citet{2014AJ....147..108L}, for each of the individual unWISE frames, a constant sky background was estimated -- as a mode of the pixel histogram --  and taken out. The fraction of pixels  containing  protocluster signal varies with the position of the protocluster and its brightness, but it is conceivable   that the background level may have been overestimated and thus oversubtracted. Since our analyses make use of the final unWISE coadds, it is not possible for us to correct individual WISE exposures for this effect nor can we track down the exact cause. 

However, this oversubtraction is  unlikely to affect our ability to measure the surface brightness profile or the total flux originating from protoclusters. Because a {\it constant} sky level is subtracted from each  tile (during the unWISE processing as well as our own), the shape of the profile on individual protoclusters is preserved  while the background level is likely set  below the true value. Thus, if we set our `sky' baseline at this level, it is possible to recover the true {\it total} flux. In  practice, this is readily achieved in all bands by requiring that, at radii larger than the expected protocluster size, the surface brightness profile asymptotes to zero, or alternatively, the cumulative flux plateaus to a constant value. 

Following \citet{2021MNRAS.501.1970A}, we create stacks from random positions to test the robustness of the total light cluster stacking method. For each of the protoclusters, we pick a random position within the same unWISE tile (1.5$^\circ$ on a side) up to 0.7$^\circ$ away from its center in both X and Y directions and perform image stacking using the identical procedure as before. The result is shown in Figure~\ref{stack_fig}.  While these ``off-cluster'' stacks indicate no detection at their centers, the noise properties are reassuringly similar to those of the protocluster stacks. For each band, we create 500 off-cluster stacks and measure the mode of the pixel distribution from each image as an estimate of the background value and its standard deviation, $\sigma_{\rm sys}$, is recorded.

\subsection{Herschel/SPIRE Image Processing}

Prior to stacking, we do not perform any additional processing of the SPIRE images. The pipeline-processed images are already background subtracted and have a mean of zero. We do, however, account for background over-subtraction after stacking as described in Appendix~\ref{appendix a}. This over-subtraction is due to the large relative size of the protoclusters to the images. 

Given the SPIRE confusion limit, the sources detected individually in the SPIRE images have flux densities $\gtrsim 20$~mJy. The number of low-redshift ($z<2$) sources at this FIR flux level is low \citep{2012A&A...542A..58B}; in turn, a significant fraction of the detected SPIRE sources are expected to lie at $z\gtrsim 2$ and possibly belong to PC15 protoclusters. For this reason, we do not mask any SPIRE detection prior to our stacking analysis.

Additionally, at these wavelength bands, the images are less dominated by foreground galaxies. \cite{2021MNRAS.501.1970A} reported that field galaxies do not contribute to the stacked signal in the SPIRE bands by stacking cutouts that were offset from the clusters by a random amount.  
We are unable to perform a similar test as the SPIRE data we do not provide a sufficiently large areal coverage to produce `off-protocluster' stacks.

\subsection{Herschel/SPIRE Stacking}

We stack the 211 protocluster candidates in the three SPIRE bands, by constructing a 3D datacube that is the size of the smallest available image dimensions for each band. The cubes are 26.5\arcmin$\times$24.0\arcmin\ for the $\mathrm{250\ \mu m}$ band, 26.6\arcmin$\times$24.2\arcmin\ for the $\mathrm{350\ \mu m}$ band, and 26.8\arcmin$\times$24.0\arcmin\ for the $\mathrm{500\ \mu m}$ band. We randomly rotate the individual cutouts in 90\arcdeg\ steps before placing them in the data cube in order to avoid systematic effects from the mapping since the Herschel scan pattern can result in offsets in the stack centers \citep{2021MNRAS.501.1970A}. We then take the inverse variance weighted mean across each pixel of the cutouts.

We compute the root-mean-square (RMS) fluctuation of the pixel distribution for each image. The average RMS over each protocluster cutout ranges from 9.3--10.9~mJy/beam at 250 $\mu m$. We do not perform sigma clipping to remove outliers because the individual noise maps are smooth and do not vary significantly.

We calculate the stacked emission in each pixel according to:

\begin{equation}
\label{eq:stack}
S_{i,j} = \sum_k\frac{I_{i,j,k}}{\sigma_{i,j,k}^2} \bigg/ \sum_k\frac{1}{\sigma_{i,j,k}^2}
\end{equation}
where $S_{i,j}$ is a pixel in the stacked image, ${I_{i,j,k}}$ is a pixel in each cutout image, ${\sigma_{i,j,k}}$ is the error for each pixel of each cutout image. We sum over the running index $k$, which is the number of image cutouts.

In Figure \ref{stack_fig} we show the protocluster stacks for the SPIRE 250, 350, and 500 $\mu$m bands. The protocluster signal is clearly detected in all bands and extends to a radius of at least 6\arcmin\ (9.0--12.5~cMpc at $z=2-4$). 

\subsection{Photometry of WISE and Herschel Stacked Images}

\label{photometry}

We perform aperture photometry on the cluster stacks as follows. In all bands, we make the cumulative flux and the surface brightness measurements in a series of circular apertures and annuli with a step size that is equal  to the beam FWHM for each band. We assume the center of the image stack to be the center of the targeted SPIRE observations. For the WISE stacks, we subtract the image by a constant background estimated at 10\arcmin--15\arcmin\ away from the image center. For the SPIRE stacks we add to the stacked image the flux that was oversubtracted as described in Appendix \ref{appendix a}. 

The cumulative flux is a sum of all pixel values within the given aperture while the radial surface brightness is measured as the mean of all pixels within a given annulus. For the SPIRE bands, the photometric uncertainties are estimated from 1,000 stacks created by bootstrapping in which a random subset of protoclusters are selected and stacked. For the WISE bands, we add the background uncertainties calculated in Section \ref{subsec:wise_stacking} to the total uncertainty.

We compare the total protocluster flux measured in the SPIRE 350~$\mu$m and SPIRE 500~$\mu$m bands to the unresolved fluxes measured by Planck for a subset of 78 PC15 sources within the PHz catalog. The average Planck fluxes of a subset of the protoclusters have larger uncertainties but are consistent with stacked SPIRE signal.

We also compare the total flux of the stacked PC15 sample to the average flux from individually detected sources in the SPIRE images. We find that on average $\sim45$\% of the total 250 $\mu$m stacked flux is from individually detected sources, although some fraction of the flux is expected to be from line of sight interlopers (see Section \ref{subsec:los}).

\subsection{Redshift Subsamples}

\label{sec:create_subsamples}

The PC15 protoclusters are expected to span  a wide redshift range of $z$=2--4. We subdivide the sample by redshift based on the flux ratios of the total SPIRE flux in different bands within a 7\arcmin\ aperture. We calculate the ratios of the total flux $F_{500}/F_{350}$ and $F_{250}/F_{350}$ as done in \cite{2015A&A...582A..30P}. Seven protocluster candidates are removed due to being individually undetected in one of the SPIRE bands.

At higher redshift, the far-IR SED peak shifts to longer wavelengths, and as a result, galaxies are expected to have a higher $F_{500}/F_{350}$ ratio and a lower $F_{250}/F_{350}$ ratio than the lower-redshift counterparts. We define our higher-$z$ galaxy subsample as those with colors $F_{500}/F_{350} > F_{250}/F_{350}$. The remainder (with colors $F_{500}/F_{350} < F_{250}/F_{350}$) are selected as the lower-$z$ subsample. This criterion results in 57 and 147 protocluster candidates in the higher- and lower-$z$ bins, respectively. We calculate the fluxes in each of the bands for the subsample stacks and include them in Table~\ref{tab:fluxes}.

\section{Correction for Line of Sight Interlopers}
\label{subsec:los}

Sources selected by instruments with low resolution, such as single-dish far-infrared instruments, suffer from contamination and subsequent flux boosting by line of sight interlopers \citep{hodge13}. Indeed, \citet{2017MNRAS.470.2253N} showed that the number density of PC15 protocluster candidates is much higher than the theoretical expectation; the implication is that a large fraction of PC15 protocluster  candidates may be superpositions of multiple,  unassociated protoclusters along the sightline. \cite{2022A&A...664A.155G} further measures the contamination along the line of sight within PC15 protoclusters in the IllustrisTNG simulation. While the manner in which star formation occurs in dense protoclusters is not well understood \citep{lim21} and may vary greatly from the expectation of dusty star formation in an average-density environment, these considerations stress the importance of properly accounting for this effect and of correcting  our measurements.

In order to realistically simulate the largest cosmic structures at high redshift, we use the halo catalog  from the WebSky Simulation \citep{2020JCAP...10..012S}. Websky is created to provide accurate extragalactic mocks to be compared against the current and next-generation cosmic microwave background observations covering large areas of the sky. 
By providing $\approx 9\times 10^8$ halos in the entire sky with the minimum halo mass  $1.2\times 10^{12}M_\odot$ out to $z=4$, the Websky catalog can account for the large-scale density fluctuations, and thus is ideal to model the all-sky Planck data and the selection of massive protoclusters therein. 

To assign an infrared luminosity to each halo, we use the $L_{\rm IR}-M_{h}-z$ calibration empirically obtained by \citet [][Equation~32]{2018MNRAS.475.3974W} assuming a 0.25~dex scatter. We model the redshift selection function by taking the $N(z)$ predictions presented in Figure~13 of \cite{2015A&A...582A..30P} and fitting them into a smooth curve with the Equation~7 of \cite{schmidt15}. These models reflect the Planck color selection corresponding to five dust temperature scenarios ranging from 25--45~K and dust emissivity $\beta=1.5$. While we take the 35~K case as our fiducial model, we compute our results in all cases.

\begin{figure}
\centering
\includegraphics[width=3.3in]{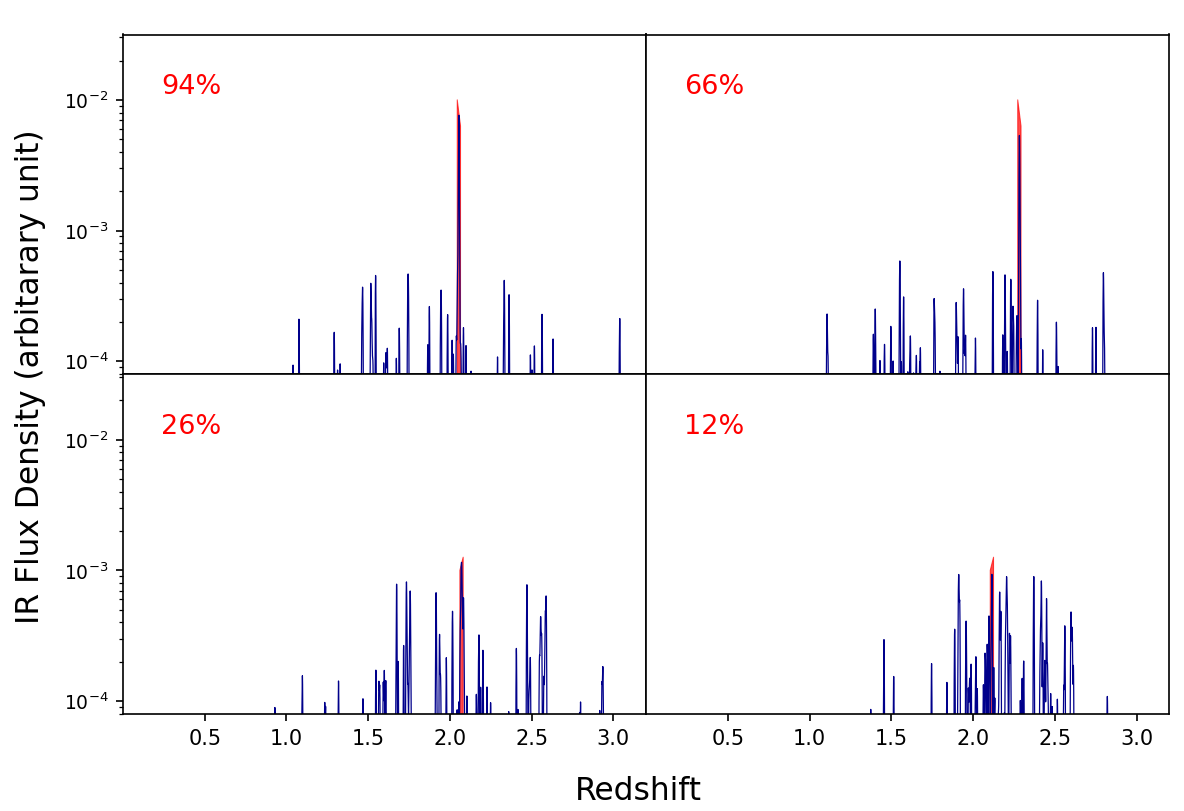}
\caption{IR flux distribution of cosmic structures constructed from the WebSky simulation are shown along four Planck-beam-sized sightlines. Flux densities of individual structures are indicated in blue while the dominant structure with the highest flux density is marked in red with its fractional contribution indicated at the top left corner of each panel. The top panels show the two brightest sources in the simulation while the bottom panels show two randomly drawn from the sample of simulated PC15 protoclusters (see Section~\ref{subsec:los} for more detail).}
\label{fig:websky_protocluster}
\end{figure}

All halo positions are projected onto the entire sky and their fluxes are pixelated by matching to the Planck beamsize of 4.5\arcmin. The 8,251 brightest IR sources are then selected above a flux density threshold, determined to satisfy the surface density of PC15 protocluster  candidates of 0.2~deg$^{-2}$ \citep{2016A&A...596A.100P}. For each of these sources, we  construct the IR luminosity contributed by different halos as a function of redshift. In Figure~\ref{fig:websky_protocluster}, we illustrate example sightlines: the top two panels show the sightlines in which  a single structure makes the highest contribution to the total flux. The bottom panels show two randomly chosen structures selected as mock 'PC15 protocluster candidates'. 

We do not directly compare the IR luminosities of these simulated PC15 protoclusters with those of the data. WebSky lacks the mass resolution to properly model star formation in halos or to identify all halos. Additionally, observed star formation in protoclusters cannot be reproduced in higher-resolution hydrodynamical simulations \citep[e.g.][]{lim21}. Regardless, the abundance matching technique we employ in this work should offer a robust way to identify structures comparable to the PC15 protoclusters provided that halo mass and IR luminosity are broadly correlated as observed \citep{2018MNRAS.475.3974W}.

The fractional contribution from the dominant structure in each sightline is computed as $f=F_1/(F_{\rm tot}-B)$ where $B$ is the background signal, $F_{\rm tot}$ is the total flux density summed over the redshift selection function, and $F_1$ is the flux density from the dominant structure. The background signal is measured by averaging over all Planck-beam-sized pixels. This step of taking out the background signal is appropriate as it mirrors the sky subtraction procedure adopted in our observational measurement (Section~\ref{subsec:wise_stacking}). 

This analysis finds, on average, that the contribution from a single protocluster in a Planck-sized beam is $33\pm 15$\%, i.e., about one third of the total flux. This is in broad agreement with \citet{2017MNRAS.470.2253N} in that the interloper contribution is non-negligible and \cite{2022MNRAS.514.5004L}, who find that almost all simulated protocluster candidates have at least two physically unassociated structures along the line of sight. Our analysis is relatively insensitive to the assumed $T_{\rm dust}$ and scatter in the $L_{\rm IR}-M_h$ relation. Assuming $T_{\rm dust}$=25, 35, and 45~K while keeping the scatter to 0.25~dex, $f$ changes from $40\pm 19$, $33\pm 15$, and $34\pm 16$\%, respectively. Similarly, changing the scatter to 0.15, 0.25, and 0.35~dex results in $30\pm 12$, $33\pm 15$, and $41\pm 22$\%, respectively. Assuming warmer dust leads to lower interloper fractions while a higher scatter increases the interloper fraction. In all cases, the change is minor compared to the intrinsic spread in the fraction of emission originating from Poisson fluctuations.

\section{Results}
\label{sec:res}

\subsection{Total Cluster Light: Radial Surface Brightness  Profiles and Fluxes of Planck Protoclusters}\label{subsec:sb_profile}
\label{sec:cigale}

\begin{figure*}[ht!]
\centering
\includegraphics[width=6in]{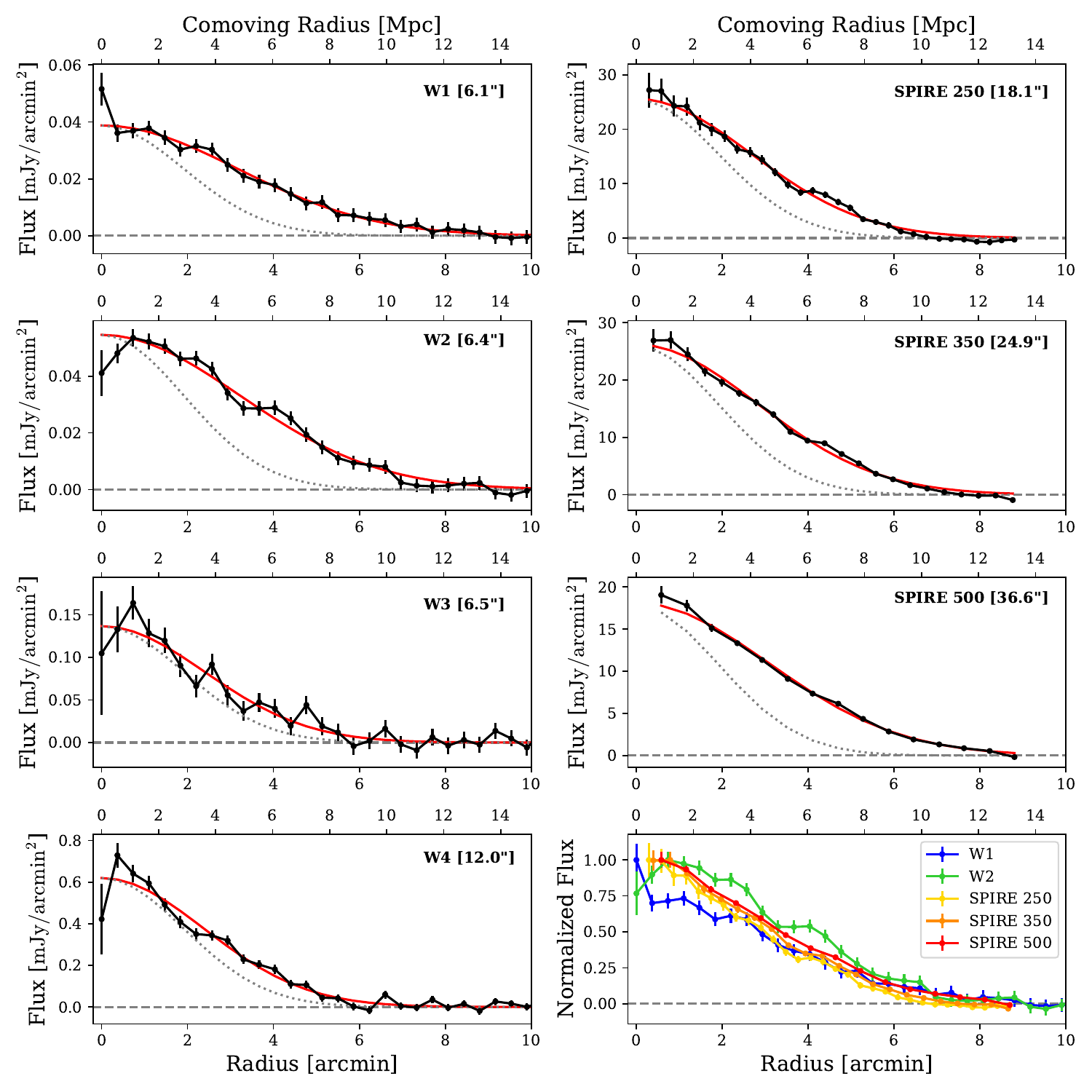}
\caption{Radial surface brightness profiles of the WISE and SPIRE stacks of the PC15 protoclusters, with the best-fit Gaussian model plotted in red. The dotted gray line indicates the profile of the Planck beam, normalized to the peak of the best-fit Gaussian model. The FWHM of the beam in each band is indicated in brackets. The bottom right panel shows the surface brightness profiles normalized to the peak value of the bands which trace stellar mass ($W1$ and $W2$), and dust-obscured star formation (SPIRE 250, SPIRE 350, and SPIRE 500). The comoving (top x-axis) scale is calculated for $z=2$.}
\label{radial_plot_fig}
\end{figure*}

In Figure~\ref{radial_plot_fig}, we show the radial surface brightness profiles of the protoclusters. 
In all bands, the shapes of the profiles are similar and extend out to 6\arcmin-8\arcmin\ from the center.  The Planck beamsize of the shortest wavelength band detection band (4.5\arcmin\ at 875 GHz) is indicated as a dotted line in each panel. We compare the radial surface brightness profiles to the Planck beam to verify that the emission is more extended than the Planck beam and therefore resolved.

Some of the WISE radial profiles show dips within the innermost region ($r<1$\arcmin). This is because our annular bins are centered on the image center (i.e., the center of the Planck detections) for consistency. Given the large beamsize of Planck, we expect that the center of each protocluster may be offset within the limit, and as a result, the stacked signal may be slightly offset as well (see Appendix~\ref{appendix b}). In addition, we note that the measurements from the innermost annular bins contain a much smaller number of pixels than measurements from the outer bins, and thus are noisier as reflected by the associated error bars. 

To measure total cluster light at each band, we perform aperture photometry of the image stack using a circular aperture with a radius of 7\arcmin.  This corresponds to 9.0--12.5~cMpc at $z=2-4$, much larger than the expected size of protoclusters \citep{2013ApJ...779..127C}. As mentioned in Section \ref{sec:method}, the surface brightness profiles level off at zero for most bands. 

The stacked total protocluster fluxes and associated uncertainties are listed in Table \ref{tab:fluxes}. It is worth noting that the SPIRE fluxes in the table are made on the full sample of 211 PC15 protocluster  candidates while the WISE fluxes are made on 152 protoclusters  (see Section~\ref{subsec:wise_stacking}) after removing 59 frames to allow clean stacking. We repeat our Herschel stacking and flux measurements on the 152 systems and find that doing so does not change the stacked average flux within the errors for all three SPIRE bands. This is not surprising considering that the removal of a given protocluster is determined by the WISE tiling scheme and the positions of bright stars in and around the WISE images  and not by the properties of protoclusters. 

We assume the Herschel observation centers on the most massive halo within (an asymmetrical) protocluster.  Since we are stacking, the random asymmetries of individual protoclusters are then washed out. However, because of the positional uncertainties of protoclusters due to the large Planck beam, the average radial profiles of the protoclusters in the stack smear out. In Appendix \ref{appendix b}, we quantify this effect by approximating the protoclusters as 2D gaussian distributions and shifting the centers. Our result shows that the effect of positional uncertainty is small.

\begin{deluxetable*}{ccccccccc}
\tablecolumns{9}
\tablewidth{0pt}
\tablehead{Sample & $N_{\rm stack}$ &            W1 &            W2 &            W3 &             W4 &         SPIRE 250 &         SPIRE 350 &        SPIRE 500}
\startdata
\input{cumulative_flux_subsets}

\enddata
\caption{Total stacked fluxes of the PC15 protoclusters in units of mJy within a circular aperture with a radius 7\arcmin\ (3.5 cMpc at $z=2$). For the higher-$z$ sample, we do not detect  any flux in $W3$ and list the 3$\sigma$ upper limit. Uncertainties are calculated by bootstrapping. The fluxes presented in this table are not corrected for line of sight interlopers (see Section \ref{subsec:los}).}
\label{tab:fluxes}
\end{deluxetable*}

\subsection{Spectral Energy Distribution}

\label{sec:sed}

\begin{figure*}[ht!]
\centering
\includegraphics[width=6in]{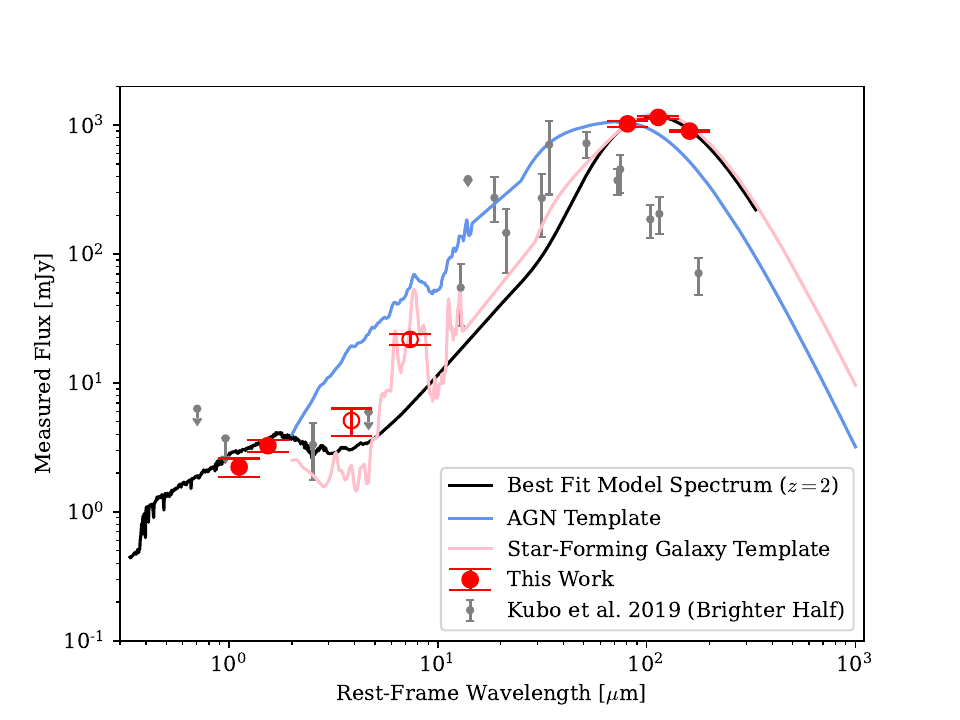}
\caption{The rest-frame SED of the full sample of stacked PC15 protocluster candidates and the best-fit model of the spectrum to the SED from CIGALE (assuming $z=2$). The data points used to fit the SED are shown by filled red circles, while the open red circles show the W3 and W4 points which are not used in the fit. AGN and star-forming galaxy templates, AGN4 and SFG3, from the comprehensive library of \cite{2015ApJ...814....9K} are plotted and normalized to the best-fit model at 83 $\mu$m (250 $\mu$m observed frame). The SED of the stack of the brighter half of the protocluster candidates from \citep{2019ApJ...887..214K} are plotted in gray.}

\label{SED_fig}
\end{figure*}

Using the flux measurements made in WISE and  SPIRE bands, we fit the spectral energy distribution (SED) fitting to derive the average physical properties of the PC15 protoclusters. Since the interloper contamination rate estimated in Section \ref{subsec:los} is not wavelength-dependent, we first run the SED fitting using the photometry measured on the image stack and correct the derived quantities after. The $W3$ and $W4$ measurements are excluded from the SED fitting since they sample the region of Polycyclic Aromatic Hydrocarbon (PAH) emission for galaxies at $z\sim 2$; the spread of the protoclusters in redshift space will cause significant scatter in $W3$ and $W4$ due to PAH emission lines moving in and out of the band (e.g\ \citealt{2021MNRAS.501.1970A}). Though we do not use the $W3$ and $W4$ bands for the SED fitting, they strongly indicate the presence of PAH features with a strength consistent with galaxy templates as shown in Figure~\ref{SED_fig}. 

We fit the observed SED to model SEDs using CIGALE, a Bayesian SED modeling code that employs multi-wavelength energy balance \citep{2005MNRAS.360.1413B, 2019A&A...622A.103B, 2009A&A...507.1793N}. We parameterize the dust emission following \cite{2012MNRAS.425.3094C}, using the {\tt casey2012} module within CIGALE, as a single-temperature modified blackbody which traces  cold dust emission from the reprocessed light from young stars, and  a mid-IR power law which traces warmer dust emission from starbursts and/or AGN. We allow the dust temperature to vary from 20~K to 60~K while the dust emissivity index ($\mathrm{\beta = 1.5}$) and the mid-IR power-law index ($\mathrm{\alpha = 2.0}$) are fixed. 

We model the star formation history (SFH) from $z=15$ up to the assumed protocluster redshift of $z=2.0$, 2.5, and 3.0. The expectation is that a protocluster SFH should increase with time at a rate commensurate with the cosmic star formation rate density (SFRD), which \cite{2014ARA&A..52..415M}  parametrized as:
\begin{equation}
\label{eq:sfh}
\psi(z) =  \frac{0.015(1+z)^{2.3}}{1+[(1+z)/2.9]^{5.6}} \mathrm{M_{\odot}\ yr^{-1}\ Mpc^{-3}}
\end{equation}
in their Equation~15. Indeed, semi-analytic models predict that the cosmic SFRD contributed by protoclusters has a similar shape as the field counterpart \citep[see Figures~4 and 5 in][]{2017ApJ...844L..23C}. For this reason, we adopt Equation~\ref{eq:sfh} as our custom SFH in CIGALE using the {\tt sfhfromfile} module. However, we caution that the true SFH of the protoclusters can be different from Equation \ref{eq:sfh} and we use this assumption since we do not have enough photometric data points to constrain the SFH.

We use the {\tt bc03} module based on the \cite{2003MNRAS.344.1000B} stellar population synthesis model, assuming solar metallicity and the \cite{2003PASP..115..763C} IMF. We use the {\tt dustatt\_powerlaw} module with an implementation of the \cite{2000ApJ...539..718C} attenuation law with 
both young and old stellar populations. We vary the $V$ band attenuation for the young stellar population from 0 to 20  while keeping the ratio of the attenuation for the old- and young populations constant at 0.44.
We set the amplitude of the UV bump to zero and assume a power-law slope of $-0.7$. The resulting best fit CIGALE SED for the full sample assuming $z=2$ is shown in Figure~\ref{SED_fig}.

We calculate $\mathrm{SFR_{IR}}$ from the the best-fit total infrared luminosity, $\mathrm{L_{IR}}$, following the calibration of \citet{2011ApJ...737...67M}:  $\mathrm{SFR_{IR}}=\mathrm{1.5\times10^{-10}\ L_{IR}}$. Since the SFR calculation within CIGALE is sensitive to the assumed star formation history, we use an empirical calibration. Assuming $z=2$, the best-fit model for the full sample yields a stellar mass of $\mathrm{4.9\pm2.2\times 10^{12}\ }M_{\odot}$ and an SFR of $7.3\pm3.2 \times 10^3~M_{\odot}$~yr$^{-1}$ after applying the interloper correction factor of $33\pm 15$ \%. The SED parameters for the full sample derived with and without correction are listed in Table~\ref{tab:parameters}. As we describe in Section \ref{sec:wise_processing}, masking out detected WISE sources would only underestimate the stellar mass by at most 6\%.

Though it is unphysical to assume that protocluster SFH remains constant or declines exponentially with time, we repeat the SED fitting with these standard assumptions. Compared to our fiducial model, a constant SFH model leads to a decrease in  stellar mass by a factor of $\approx2$ while the best-fit exponentially decaying model results in an increase of stellar mass by up to 50\%. SFRs remain unchanged within a few percent regardless of the assumed SFHs.

\subsection{Redshift Evolution within the Sample}

\label{sec:subsamples}

We look for redshift evolution within the protocluster sample, by subdividing the full sample based on SPIRE colors as discussed in Section \ref{sec:create_subsamples}. We  fit the SEDs of the subsamples with CIGALE, using the identical setting described in Section~\ref{sec:cigale} other than the fact that we assume  $z=2$ and $z=3$ for the two subsamples. We estimate the redshifts of the subsamples based on the SPIRE color-color plot in Figure 8 from \cite{2015A&A...582A..30P}. The results of the SED fitting are presented in Table~\ref{tab:parameters}. Our results suggest that galaxies in higher-redshift protoclusters have higher specific star formation rates ($2.9 \pm 0.3\times 10^{-9}\ yr^{-1}$) than those in their lower-redshift counterparts ($1.5 \pm 0.1 \times 10^{-9}\ yr^{-1}$), consistent with the trend seen in field galaxies \citep{2014ApJS..214...15S}.

\begin{deluxetable*}{cccccccc|ccc}
\tablecolumns{11}
\tablewidth{0pt}
\tablehead{ Sample & $z$ & $\chi^2_{\nu}$ &   $\mathrm{T}$ &               $\mathrm{L_{IR}}$ &             $\mathrm{M_{\ast}}$ &                          $\mathrm{SFR}$ &                $\mathrm{sSFR}$ &         $\mathrm{L_{IR, corr}}$   &  $\mathrm{M_{\ast, corr}}$ &                   $\mathrm{SFR_{corr}}$ \\
&  &  & $\mathrm{[K]}$ & $\mathrm{[10^{14}\ L_{\odot}]}$ & $\mathrm{[10^{12}\ M_{\odot}]}$ & $\mathrm{[10^{3}\ M_{\odot}\ yr^{-1}]}$ & $\mathrm{[10^{-9}\ yr^{-1}]}$ & $\mathrm{[10^{14}\ L_{\odot}]}$ & $\mathrm{[10^{12}\ M_{\odot}]}$ & $\mathrm{[10^{3}\ M_{\odot}\ yr^{-1}]}$ 
}
\startdata
\input{final_table}

\enddata
\caption{Parameters from the best-fit SED to the PC15 protocluster stacks. We present the total stellar mass and total star formation rate corrected and not corrected for line of sight interlopers. The uncertainties in the temperature are based on the CIGALE grid spacing of 5 K.}
\label{tab:parameters}
\end{deluxetable*}

\section{Discussion}
\label{sec:disc}

\begin{figure*}[ht!]
\centering
\includegraphics[width=6in]{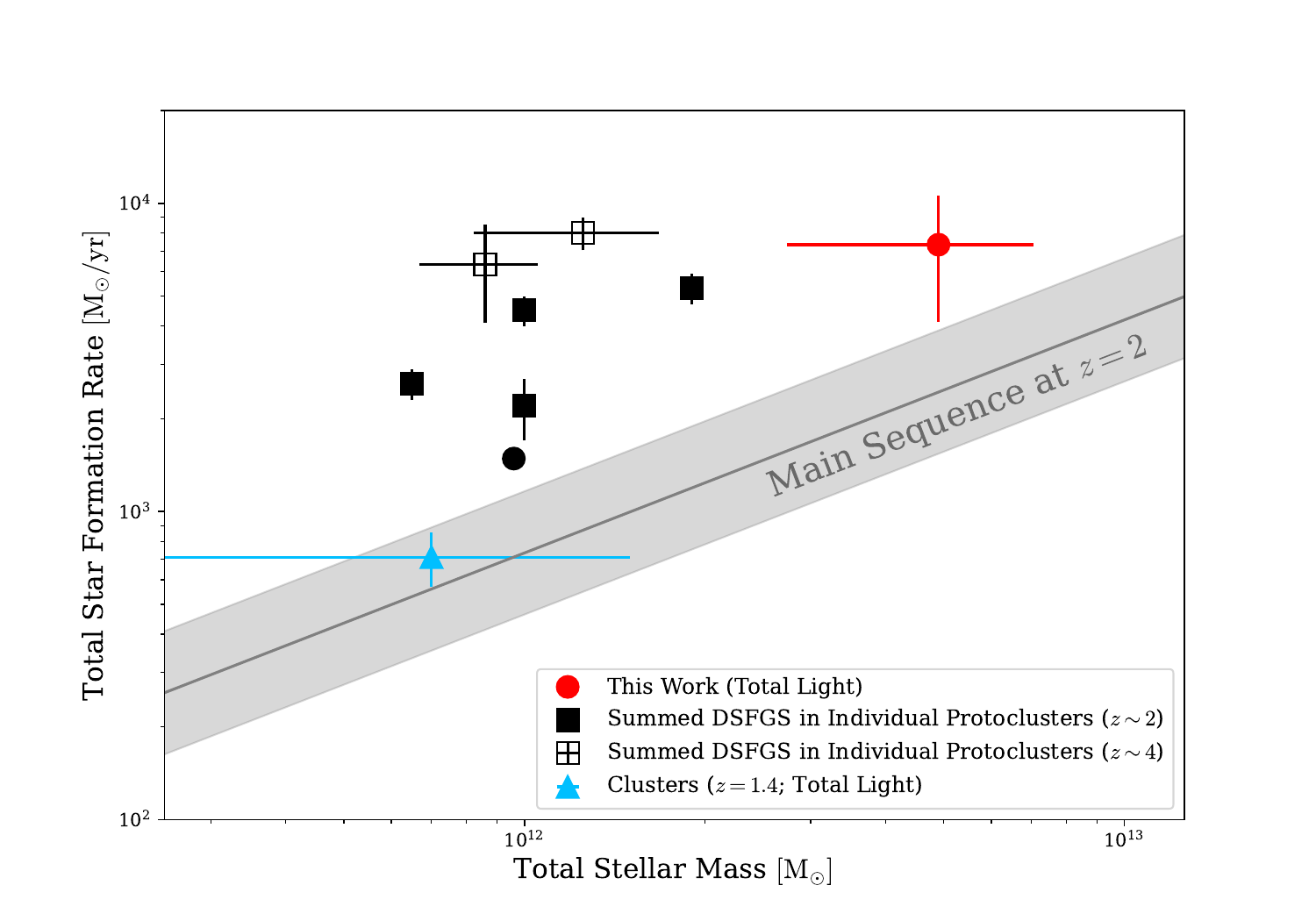}
\caption{Total star formation rates as a function of total stellar masses for different studies of protoclusters and clusters. Red indicates our total light measurement for PC15 protoclusters corrected for LOS interlopers. We also plot individual protoclusters \citep{2016ApJ...824...36C} that contain DSFGs (square points), and a confirmed Planck-selected protocluster from the PHz catalog \citep{2021A&A...654A.121P} indicated by the circular point. We also show $z\sim4$ protoclusters \citep{2020ApJ...898..133L,2020MNRAS.495.3124H,2022MNRAS.512.4352H} and the $z=1.4$ cluster sample from \cite{2021MNRAS.501.1970A}. The main sequence at $z=2$ from \cite{2014ApJS..214...15S} is extrapolated to higher stellar masses and shown in gray.}
\label{compare_proto} 
\end{figure*}

\subsection{The Spatial Extent of Stellar Light and Dusty Star Formation in Protoclusters}

We compare the normalized radial surface brightness profiles of the stacks in the WISE $W1$ and $W2$ bands and in the SPIRE bands in Figure \ref{radial_plot_fig} (bottom right panel). The WISE $W1$ and $W2$ bands trace the stellar light, while the SPIRE bands trace the dust emission. We determine the half-light radius of each of the surface brightness profiles by fitting a 1D Gaussian model to each profile and we list them in Table~\ref{tab:half_light_radii}. We measure the half-light radii to be around 2.8\arcmin\ (corresponding to 4.2--5.8~cMpc at $z=2-4$). The angular extent of the emission in all of these bands is similar, although the radial surface brightness profiles in the WISE bands appear to have a slightly larger extent. We cannot say if this is because the stellar light traces a larger area than the dust emission or if this is a systematic effect due to the lower signal-to-noise ratio of the WISE stacks. We also determine the half-light radii for the lower-$z$ and higher-$z$ subsamples and, when detected, find them to be consistent with the full sample. 

The half-light radii we measure may be overestimated due to positional uncertainty given the Planck beamsize. In Appendix~\ref{appendix b}, we simulate this effect and find that a simulated protocluster light profile has a FWHM that is on average 14\% smaller in all bands than the measured FWHM of the protocluster profiles. We list the corrected half-light radii in Table~\ref{tab:half_light_radii}. 
 
\cite{2013ApJ...779..127C} use numerical simulations to predict the extent of protoclusters at different redshifts for structures that end up with halo masses ranging from $10^{14} M_{\odot}$ to $10^{15} M_{\odot}$ at $z=0$. As we discuss in Section \ref{halo_mass}, we estimate that the PC15 protoclusters represent Coma-cluster progenitors, with halo masses greater than $\approx10^{15}M_{\odot}$. The effective radii of these simulated protoclusters correspond to the angular radii of 3\arcmin--7\arcmin, broadly consistent with our measured half-light radii.

\cite{2022MNRAS.514.5004L} also perform a stacking analysis of the PC15 protocluster candidates, although they center each image on the brightest SPIRE source prior to stacking. This will cause the radial profile to be a combination of  a bright central point source and extended emission. Ignoring the bright central point source, the \cite{2022MNRAS.514.5004L} stacks show extended emission out to a radius of around 5\arcmin\ where the radial profile flux decreases to zero, similar to our results ($\sim$7\arcmin\ as shown in Figure~\ref{radial_plot_fig}).

\begin{deluxetable*}{cccccccc}
\tablecolumns{8}
\tablewidth{0pt}
\tablehead{Sample &            W1 &            W2 &            W3 &             W4 &         SPIRE 250 &         SPIRE 350 &        SPIRE 500}
\startdata
\input{half_light_radii}
\enddata
\caption{Half-light radii from the radial surface brightness profiles in each band of the PC15 protoclusters, corrected for positional uncertainty in the Planck beam as described in Appendix \ref{appendix b}. The measured half-light radii are indicated in brackets. The uncertainties range from 0.05\arcmin\ to 0.3\arcmin. For the higher-z sub-sample, we do not detect any flux in $W3$ and do not measure a robust radial profile in $W4$.}
\label{tab:half_light_radii}
\end{deluxetable*}

\begin{figure}
\centering
\includegraphics[width=3.3in]{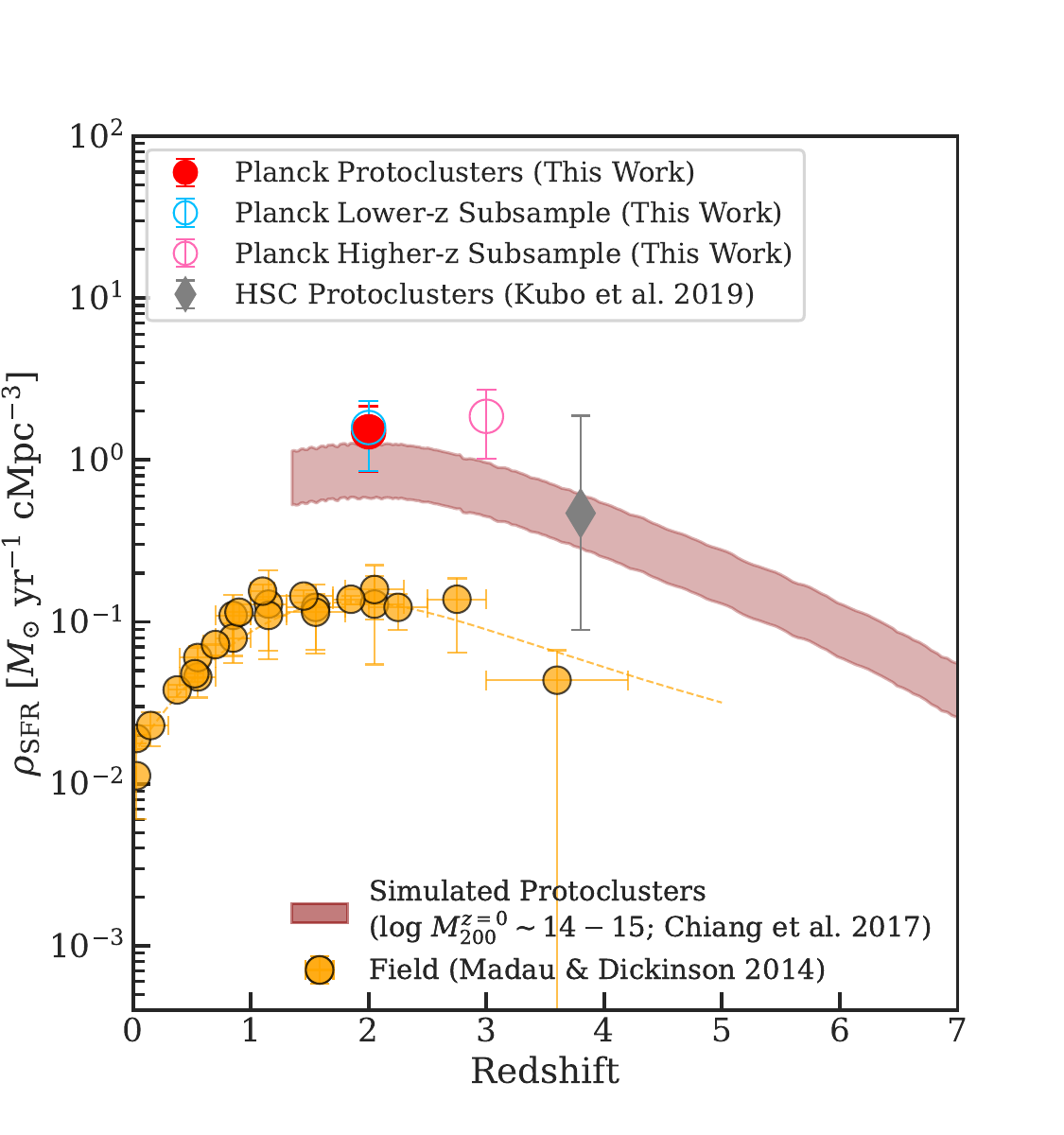}
\caption{The star formation rate density (SFRD) of the PC15 protocluster candidates, including lower-z and higher-z subsamples, and the LBG selected Hyper Suprime-Cam protocluster candidates \citep{2019ApJ...887..214K} in comparison to the projected SFRD for protoclusters with a $z = 0$ halo mass of $log\ M_{200}/M_{\odot} \sim 14-15$ scaled from \citet{2017ApJ...844L..23C} and the observational measurements of field galaxies \citep{2014ARA&A..52..415M} This figure is adapted from \cite{2022Univ....8..554A}.}.
\label{sfrd}
\end{figure}

\subsection{Total Stellar Mass and Total Star Formation in Protoclusters}

\subsubsection{Comparison to Hyper Suprime-Cam Protoclusters ($z=3.3-4.2$)}

We compare our results to the star formation rate estimated for another sample of protoclusters from \cite{2019ApJ...887..214K}, which are selected as overdensities of Lyman Break Galaxies (LBGs) identified from the Hyper Suprime-Cam Subaru Strategic Program \citep{2018PASJ...70S...4A}. The protocluster sample from \cite{2019ApJ...887..214K} is located at $z\sim3.8$, have a number density of $n = 4.6 \times 10^{-7} \mathrm{h^{3} Mpc^{-3}}$ and an average halo mass of $M \sim(0.8-1.1) \times 10^{13} \mathrm{h^{-1} M_{\odot}}$ \citep{2018PASJ...70S..12T}. \cite{2019ApJ...887..214K} employed a similar stacking technique to our analysis, which should capture the total light from the protoclusters. We compare the rest-frame average SED of the PC15 protocluster candidates to that of the protocluster candidates from \cite{2019ApJ...887..214K} in Figure \ref{SED_fig} and find that the SEDs peak at different wavelengths. The SED of the protocluster candidates from \cite{2019ApJ...887..214K} implies much hotter dust, which is inconsistent with a  purely star-formation dominated SED. \cite{2019ApJ...887..214K} fit a composite star-forming and AGN template and obtain a  total SFR of $\mathrm{2.1^{+6.3}_{-1.7}\ \times 10^3\ M_{\odot} yr^{-1}}$, which is consistent with the average star-formation rate we calculate for the  PC15 protocluster candidates, indicating that both samples have large amounts of star formation, but the \cite{2019ApJ...887..214K} sample potentially has a large AGN contribution to its dust emission.

In comparison, we do not find evidence for a strong AGN contribution within our sample. 
The theoretical AGN models from \cite{2006MNRAS.366..767F}  and \cite{2014ApJ...784...83D}  provide poor fits to our data. 
We also overplot the empirical AGN and star-forming galaxy templates from \cite{2015ApJ...814....9K} in Figure~\ref{SED_fig} and visually compare the SEDs to the templates and see that the shift in the peak to shorter wavelengths due to AGN is not seen in the average SED of the PC15 protocluster candidates, but can be seen in the average SED of the \cite{2019ApJ...887..214K} protocluster candidates. We conclude there is little evidence for a dominant AGN contribution to the average SED within the PC15 protocluster sample. The lack of warm dust inferred from our SED fitting may be a selection effect from the Planck color selection. 

\subsubsection{Fraction of Flux from Bright DSFGs}

We estimate the fraction of the total SFR and stellar mass of the PC15 protoclusters from galaxies which are undetected on the SPIRE images by comparing the total SFR and stellar mass from our total light stacking to the sum of the SFR and stellar mass from spectroscopically-confirmed bright DSFGs in protoclusters: four protoclusters at $z\sim2$ \citep{2016ApJ...824...36C}, one Planck selected protocluster at $z = 2.16$ \citep{2021A&A...654A.121P} and two protoclusters at $z\sim4$ \citep{2020ApJ...898..133L,2020MNRAS.495.3124H,2022MNRAS.512.4352H}. The individual protoclusters have halo masses greater than $10^{13}\ M_{\odot}$ at their respective redshifts, which places them on similar evolutionary tracks as the PC15 protoclusters \citep{2013ApJ...779..127C}. We define low luminosity galaxies as those not detectable individually on the SPIRE images, corresponding to the flux $F_{250}< 20$ mJy at 250 $\mu$m \citep[$L_{IR} \sim 3\times 10^{12} L_{\odot}$ assuming a star-forming galaxy template at $z=2$ from][]{2012ApJ...759..139K}.

In Figure \ref{compare_proto}, we plot the total stellar masses and star formation rates calculated for the PC15 protocluster sample in comparison with summed stellar mass and star formation rate values from the literature of individual spectroscopically confirmed DSFGs in protoclusters. Our stacking analysis of the PC15 protoclusters finds two times more SFR and four times more stellar mass than the sum of the SFRs and stellar masses of sources detected individually in DSFG overdense protoclusters. This suggests that much of the light from protoclusters comes from less luminous member galaxies, which is also consistent with studies of lower-z galaxy clusters \citep{2021MNRAS.501.1970A,2022ApJ...928...88M}.

We find that around 50\% of far-IR light within our protocluster sample could come from detected DSFGs. In blank fields, \cite{2012A&A...542A..58B} reported that DSFGs with fluxes greater than 20~mJy account for only 15\% of the integrated light in the 250~$\mu$m band. The threshold of 20 mJy at 250~$\mu$m in \cite{2012A&A...542A..58B} is chosen to correspond to sources detected with $\sim3-4\sigma$, where $\sigma$ includes confusion, in SPIRE 250 $\mu$m maps. This is roughly the same limit as the submm followup of the DSFGs with $F_{250\ \mu m}\geq 20$ mJy in protoclusters. Our results indicate that the PC15 protocluster candidates have an abundance of luminous DSFGs relative to the field.

\subsection{Star Formation Rate Density of Protoclusters}

Given the total star formation that we measure for the PC15 protocluster sample, we calculate the star formation rate density (SFRD) of the PC15 protoclusters assuming that a protocluster volume is approximated by a sphere of a radius 10.5~cMpc at $z=2$ (corresponding to the 7\arcmin\ aperture in which SFR and stellar mass are measured). We include the lower-z and higher-z subsamples of the PC15 protoclusters discussed in Section \ref{sec:subsamples}, which are consistent with the same SFRD given the uncertainties.

We also compare to the LBG-selected protocluster candidates from \cite{2019ApJ...887..214K}, since this is the only other protocluster sample that has been analyzed through total light stacking. Though the LBG-selected protoclusters are not expected to be the progenitors of the PC15 protoclusters, we include them as a reference point to show the evolution of different protocluster populations.
We calculate the SFRD of the LBG-selected protocluster candidates where we assume a spherical volume with a radius of 10.2~cMpc, as determined by a radius of 5\arcmin\ at $z=3.8$.

The result is illustrated in Figure \ref{sfrd} together with the cosmic SFRD of the field \citep{2014ARA&A..52..415M}.
We also show the \citet{2017ApJ...844L..23C} predictions  based on semianalytic models, which show  the evolution of galaxy clusters at $z=0$ with halo masses greater than $10^{14}M_\odot$ (the median mass is $M_{200}=10^{14.3}M_\odot$).

Our SFRD measurement lies an order of magnitude above the field expectation. The SFRD values from both the PC15 protoclusters and the LBG-selected protoclusters are consistent with the theoretical predictions of \citet{2017ApJ...844L..23C}.

\begin{figure}
\centering
\includegraphics[width=3.3in]{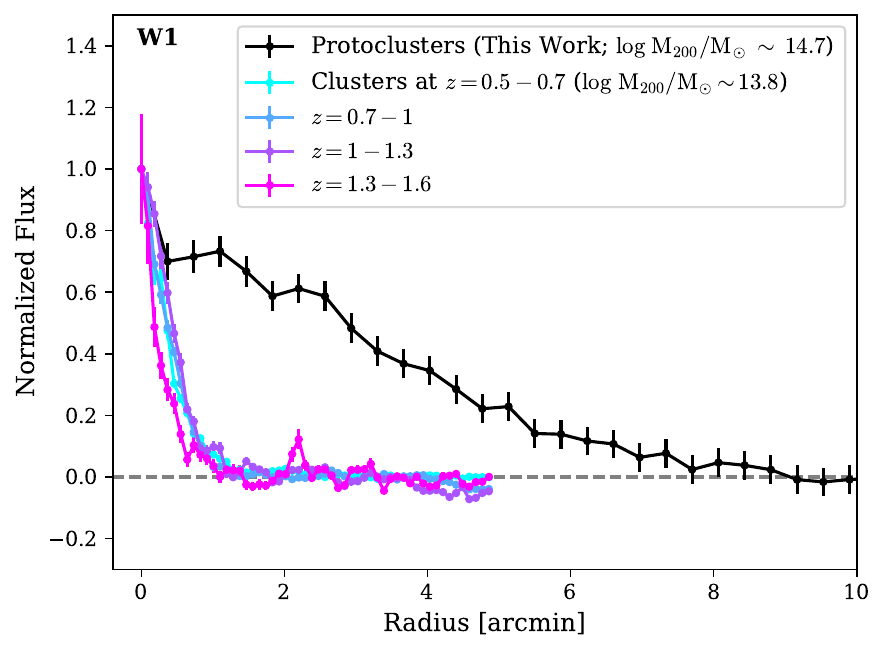}
\caption{Surface brightness profiles in the $W1$ band of the PC15 protoclusters (black) and  of the clusters studied by \cite{2021MNRAS.501.1970A} in redshift bins of $z=0.5-0.7$, $z=0.7-1.0$, $z=1.0-1.3$, and $z=1.3-1.6$ (colored lines). }
\label{radial_plot_bootes_fig}
\end{figure}

\subsection{Evolution from Protoclusters to Clusters}
\label{halo_mass}

Following the method used by \cite{2020ApJ...898..133L}, we estimate the average mass of the dark matter halos hosting the PC15 protoclusters by converting the total stellar mass to a total mass by assuming a constant stellar-to-halo-mass ratio, $\mathrm{M_*/M_{halo}}$.  

Motivated by \citet[][see their Figure~7b]{2013ApJ...770...57B}, we adopt $\mathrm{M_*/M_{halo}} = 0.01$, representative of $\approx10^{11}M_\odot$ halos hosting galaxies with stellar masses $\approx10^{9}M_\odot$ at $z\sim2$. Recent surveys showed that the galaxy stellar mass function increases steeply towards the low-mass end \citep[with $\alpha\approx -1.6$: e.g.,][]{2022ApJ...940..135S}; as a result, the stellar mass budget is dominated by low-mass galaxies.
We then calculate the total halo mass of PC15 protoclusters to be $\mathrm{4.9\pm2.2\times 10^{14}\ M_{\odot}}$.

This estimate is uncertain as it sensitively depends on the adopted $\mathrm{M_*/M_{halo}}$ value and the line of sight interloper correction. 
The former is known to vary with a galaxy's stellar mass \citep{2013ApJ...770...57B} while the latter is based on the empirical `field' expectation of how halo mass correlates with infrared luminosity  (see Section~\ref{subsec:los}). It is also conceivable that both relations may scale differently in protocluster environments. 
Still, the total halo mass of PC15 protoclusters is  consistent with that of Coma progenitors at $z=2-4$  \citep{2013ApJ...779..127C}, reinforcing the notion that they represent the most massive cosmic structures in the universe. 

We compare the spatial extents of the PC15 protoclusters to those obtained by the stacked observations of galaxy clusters ($\mathrm{log ~M_{200}/M_{\odot} \sim 13.8}$) in the Bo{\"o}tes field from \cite{2021MNRAS.501.1970A}. The halo mass of the PC15 protoclusters is estimated to be much greater than the halo mass determined for the Bo{\"o}tes clusters. The most massive of the stacked Bo{\"o}tes clusters have average stellar masses around $\mathrm{1.4 \times 10^{12}\ M_{\odot}}$, while the PC15 protoclusters have an average stellar mass around $\mathrm{4.7 \times 10^{12}\ M_{\odot}}$. As such, the PC15 protoclusters are much more massive than the Bo{\"o}tes clusters and are unlikely to be the progenitors of the Bo{\"o}tes clusters. However, given that this is the only cluster sample with a measurement of the total cluster light we compare the profile of the PC15 protoclusters to the Bo{\"o}tes clusters.

The Bo{\"o}tes clusters were split into four redshift bins: $z=0.5-0.7$, $z=0.7-1.0$, $z=1.0-1.3$, and $z=1.3-1.6$ and were stacked in both WISE and SPIRE using the same method that we implement. In Figure \ref{radial_plot_bootes_fig} we overplot the normalized radial surface brightness profiles of the Bo{\"o}tes clusters on top of the normalized radial profile of the protoclusters in $W1$. The Bo{\"o}tes clusters in all redshift bins have half-light radii around 0.5\arcmin\ (0.5 cMpc at $z=1.15$), while the protoclusters have a half-light radius of around 3\arcmin\ (4.5 cMpc at $z=2$). The extent of the Bo{\"o}tes clusters is consistent with their virial radii. Though the PC15 protoclusters are not expected to be progenitors of the Bo{\"o}tes clusters, the difference in radial extent is consistent with theoretical predictions from \cite{2013ApJ...779..127C}, which suggest that protoclusters decrease in angular size as traced by both stellar mass and star formation as they assemble and gravitationally coalesce.

\section{Conclusions}
\label{sec:conclusion}

We study the physical properties of $z=2-4$ protocluster candidates identified by \cite[][PC15 protoclusters]{2015A&A...582A..30P}. We perform a stacking analysis of WISE and {\it Herschel} images of the PC15 protoclusters to study the average properties of {\it all} galaxies residing in these cosmic structures. Total cluster light stacking analyses performed in this work complement the existing deep spectroscopic observations of individual protoclusters which exist only for a handful of systems.
 
Our main conclusions are as follows:

\begin{enumerate}
    \item  The PC15 protoclusters have a similar extent in the WISE $W1$ and $W2$ bands and the three SPIRE bands indicating that the stellar light and the light from dust trace similar regions. 
   
    \item We fit the SED of the PC15 protocluster stacks and determine the total star formation rate and the total stellar mass. When compared to the total star formation from individual DSFGs in protoclusters we find a total SFR that is two times larger and a stellar mass that is four times larger, indicating that much of the star formation and stellar mass in protoclusters comes from galaxies with lower luminosities than DSFGs. 

    \item From the SED fits of the PC15 protoclusters, we do not see any evidence of additional warm dust that would indicate a significant AGN contribution, contrary to the results of \citep{2019ApJ...887..214K}. This could be due to the different methods for selecting these respective protocluster samples.

    \item  We measure the half light radii of the protoclusters to be $\sim2.8$\arcmin\ (corresponding to 4.2--5.8~cMpc at $z=2-4$). This angular size is consistent with the evolution of Coma-cluster progenitors from \cite{2013ApJ...779..127C}. The PC15 protoclusters have a more extended radial distribution than the stacked measurement of low-redshift clusters \citep{2021MNRAS.501.1970A}, which is in qualitative agreement with the expectation that protoclusters are not yet collapsed.
    
\end{enumerate}

The total light stacking analysis that we present can be applied to other samples of protoclusters and clusters selected by different methods.  Follow-up with upcoming mm-wavelength surveys by TolTEC on the Large Millimeter Telescope will reveal the amount of dust-obscured star formation within protocluster candidates. Upcoming large wide-field surveys such as the Vera C. Rubin Observatory's Legacy Survey of Space and Time, Euclid, and the Nancy Grace Roman Space Telescope will generate large samples of candidate protoclusters.  A stacking analysis can compare the different populations of protocluster candidates selected by different methods, determine how much star formation there is within populations of protoclusters, and determine where the cosmic star-formation rate density of protoclusters peaks.

\begin{acknowledgments}

The authors thank the referee for a thorough review of this paper which improved the clarity of the work. The authors thank H. Dole for providing the list of coordinates of the PC15 protocluster candidates. The authors also thank M. Boquien and D. Burgarella for advice for using CIGALE, as well as D. Lang and A. Meisner for assistance with downloading unWISE images.  Support for the work was provided by
NASA through the Astrophysics Data Analysis Program, grant number 80NSSC19K0582 and through the Massachusetts Space Grant Consortium (MASGC). S.A. acknowledges support from the James Webb Space Telescope (JWST) Mid-Infrared Instrument (MIRI) Science
Team Lead, grant 80NSSC18K0555, from NASA Goddard Space Flight Center to the University of Arizona. Y.C. acknowledges the support of the National Science and Technology Council of Taiwan through grant NSTC 111-2112-M-001-090-MY3. This publication makes use of data products from the Wide-field Infrared Survey Explorer, which is a joint project of the University of California, Los Angeles, and the Jet Propulsion Laboratory/California Institute of Technology, and NEOWISE, which is a project of the Jet Propulsion Laboratory/California Institute of Technology. WISE and NEOWISE are funded by the National Aeronautics and Space Administration. The Herschel spacecraft was designed, built, tested, and launched under a contract to ESA managed by the Herschel/Planck Project team by an industrial consortium under the overall responsibility of the prime contractor Thales Alenia Space (Cannes), and including Astrium (Friedrichshafen) responsible for the payload module and for system testing at spacecraft level, Thales Alenia Space (Turin) responsible for the service module, and Astrium (Toulouse) responsible for the telescope, with in excess of a hundred subcontractors.

\end{acknowledgments}

\bibliography{main}
\bibliographystyle{aasjournal}

\appendix

\section{Accounting for Herschel Background Oversubtraction}

\label{appendix a}

\begin{figure*}[!ht]
\centering
\includegraphics[width=7in]{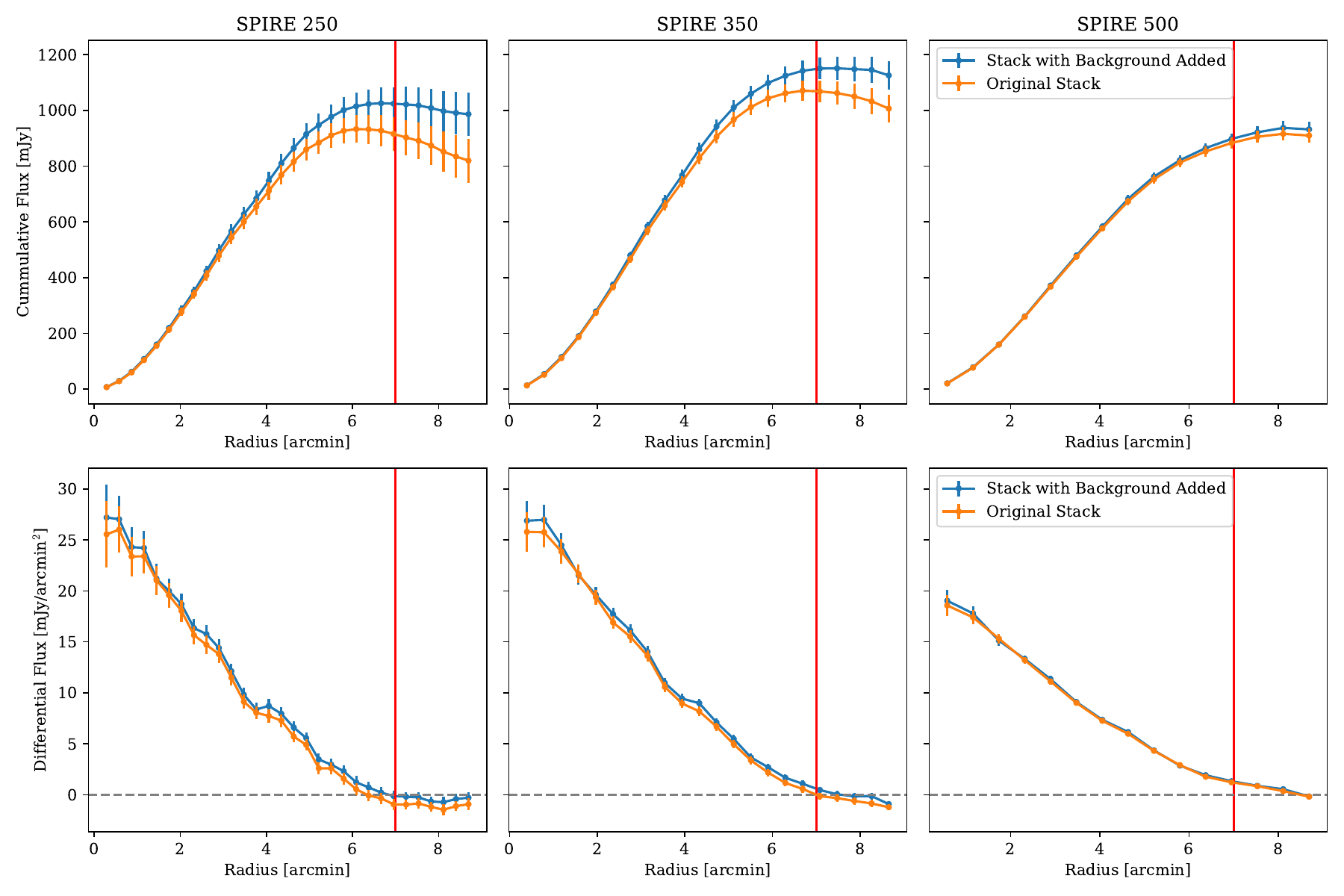}
\caption{Cumulative and differential flux profiles for  both the original stacked images (orange) and for those with background over-subtraction added  back in (blue). The red vertical line indicates 7\arcmin,  the size of the aperture used to measure the total  flux.
}
\label{background_added}
\end{figure*}

The SPIRE data reduction pipeline subtracts the background from the images such that the images have zero mean background. As described in Section \ref{sec:data}, the images can be as small as 24\arcmin\ across while we calculate that the protoclusters can extend to have diameters of around 14\arcmin, where the radial profile flux decreases to zero. Due to the large relative size of the protoclusters to the images, there could be a problem with the automatic background subtraction such that it over-subtracts. This is observed in the cumulative flux profiles that do not level off to a constant value at large radii, and the differential flux profiles which go below zero as shown in Figure \ref{background_added}. We calculate this over-subtraction by calculating the average of the radial profile bins that have differential fluxes below zero. We add this small value back to the stacks: 0.007 mJy/pix, 0.014 mJy/pix, and 0.005 mJy/pix to the images in the 250, 350, and 500~$\mu$m bands respectively. We estimate the error from this background addition to the cumulative flux within an aperture of 7\arcmin, which is the size of the aperture we use to calculate the total flux measurements.
We calculate errors of 26.8~mJy, 23.4~mJy, and 27.8~mJy for each of the 250, 350, and 500~$\mu$m bands. For the 250 and 350~$\mu$m bands, these errors are small compared to the total errors, which are 59.9~mJy and 37.8~mJy, respectively. For the 500~$\mu$m band, the error from the background correction is of the same order as the bootstrap error of 19.9~mJy. Since the signal-to-noise ratio of the SPIRE measurements is so high, adding this additional error will have a negligible impact so we do not add these errors into the flux measurements. For the lower-z  and higher-z subset sample stacks we apply the same background values that we determined for the full sample stacks.

\section{Effect of Planck Beamsize on Measured Protocluster Size}

\label{appendix b}

In determining the angular size of the Planck protoclusters, the main limiting factor is the positional uncertainty within the beamsize of the Planck High Frequency Instrument \citep[][]{2015A&A...582A..30P}, which is 4.5\arcmin\ in the detection band. This means that the sources that contribute to the Planck flux, protoclusters and interlopers alike, are randomly offset in position relative to the center of the Planck beam and with one another. Thus, the measured angular profile of the  protocluster stack is expected to be broadened by this uncertainty. 

To quantify the extent of this  effect and correct for it,

we create simulated protocluster stacks as follows. We place 211 mock protoclusters on a two-dimensional image where each source is modeled as a 2D Gaussian with a fixed FWHM while the source position relative to the image center is determined at random following a normal distribution with the width of the Planck beam. Ten input FWHM values are used ranging from 3.5\arcmin\ to 8.0\arcmin\ in increments of 0.5\arcmin. We then resample the image at the pixel scale of our data and convolve it with the PSF of each band to create a mock protocluster stack. The FWHM of the profile is measured using {\tt scipy.curve\_fit} assuming a Gaussian model.

\begin{figure*}[!ht]
\centering
\includegraphics[width=7in]{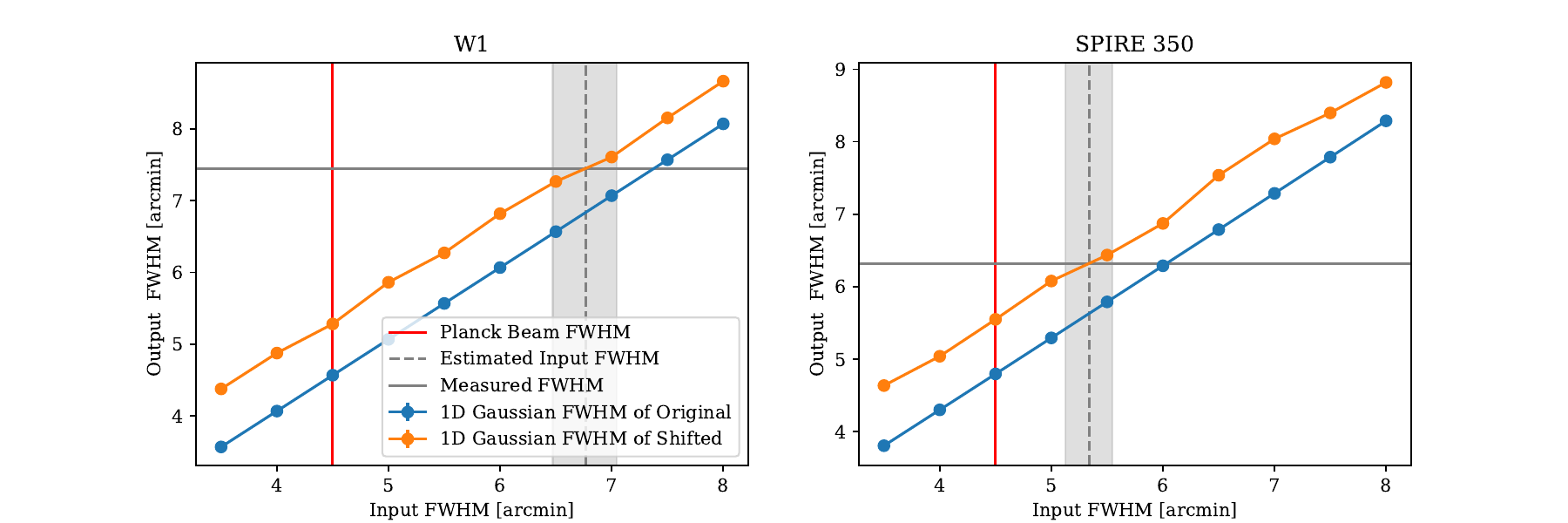}
\caption{Input vs measured FWHMs of simulated protocluster stacks are shown in orange. A one-one relation is shown in blue while the red vertical line indicates the FWHM of the Planck beam. The solid gray horizontal line marks the measured FWHM from our data. The vertical dashed line and the shaded region, both in gray, show the range of the {\it inferred} protocluster size after correcting for the broadening effect.}
\label{pos_uncert}
\end{figure*}

In Figure \ref{pos_uncert}, we illustrate the result of our simulation for the unWISE $W1$ and Herschel SPIRE $350$~$\mu$m bands in orange. The one-to-one relation is shown in blue and the Planck beam FWHM is indicated by the red vertical line. The horizontal grey line indicates the protocluster FWHM {\it as measured} while the vertical dashed line and the swath, both in grey, mark the inferred FWHM of our protocluster after the correction. In all cases, the protocluster emission is clearly resolved (i.e., more extended than the Planck beam). We repeat the similar analyses in all unWISE and Herschel SPIRE bands and find that the level of broadening is 

10\%, 9\%, 18\%, and 21\% for the $W1$, $W2$, $W3$, $W4$ and 18\%, 16\%, and 18\% for the SPIRE 250,  350 and  500~$\mu$m bands, respectively. 

In Section~\ref{subsec:los}, we find that only $\approx$one third of the measured flux belongs to the `main' protocluster while the remaining $\approx$two thirds come from interlopers which themselves may be (typically unassociated) smaller structures. In light of this, we run another set of simulations intended to set the most conservative upper limit in a configuration that mimics the reality. The procedure differs from the previous simulation in two aspects: first, we place the main protocluster, represented by a Gaussian model, carrying 1/3 of the flux at image center while other sources carry the rest of the flux. Second, the sources are this time distributed {\it uniformly} within the Planck beamsize. 
This setup results in more severe broadening of up to 35\% compared to our nominal simulation (up to $\approx$20\%), as expected. By changing the fractional contribution of interlopers and protoclusters, we find that the increased level of broadening is primarily attributed to the uniform distribution of sources within the Planck beam, which is unlikely to reflect the reality.

\end{document}

%% file: cumulative_flux_subsets.tex
   Full Sample &             211 & 2.2 $\pm$ 0.4 & 3.3 $\pm$ 0.4 & 5.1 $\pm$ 1.2 & 21.8 $\pm$ 2.1 & 1024.7 $\pm$ 59.9 & 1150.0 $\pm$ 37.8 & 900.3 $\pm$ 19.9 \\
 Lower-z Sample &             147 & 2.8 $\pm$ 0.4 & 3.6 $\pm$ 0.4 & 7.9 $\pm$ 1.2 & 21.4 $\pm$ 2.1 & 1198.8 $\pm$ 63.2 & 1234.4 $\pm$ 45.5 & 928.8 $\pm$ 23.7 \\
Higher-z Sample &              57 & 0.9 $\pm$ 0.4 & 2.2 $\pm$ 0.4 &       $<$ 3.0 & 16.6 $\pm$ 2.4 &  800.7 $\pm$ 71.7 & 1034.1 $\pm$ 61.6 & 866.0 $\pm$ 32.0 

%% file: final_table.tex
        Full Sample & 2.0 &           1.4 &       $40\pm5$ &                   $1.53\pm0.08$ &                    $15.3\pm0.8$ &                            $22.9\pm1.1$ &                   $1.5\pm0.1$ &                   $0.49\pm0.22$ &                     $4.9\pm2.2$ &                             $7.3\pm3.2$ \\
                     & 2.5 &           0.8 &       $50\pm5$ &                   $2.73\pm0.15$ &                    $20.0\pm1.1$ &                            $40.9\pm2.3$ &                   $2.0\pm0.2$ &                   $0.87\pm0.41$ &                     $6.4\pm3.0$ &                            $13.1\pm6.1$ \\
                     & 3.0 &           0.5 &       $60\pm5$ &                   $4.35\pm0.26$ &                    $25.4\pm1.5$ &                            $65.3\pm3.9$ &                   $2.6\pm0.2$ &                   $1.42\pm0.63$ &                     $8.3\pm3.7$ &                            $21.3\pm9.4$ \\
 Lower-$z$ Subsample & 2.0 &           1.4 &       $40\pm5$ &                   $1.62\pm0.08$ &                    $16.3\pm0.8$ &                            $24.4\pm1.2$ &                   $1.5\pm0.1$ &                   $0.52\pm0.24$ &                     $5.2\pm2.4$ &                             $7.8\pm3.6$ \\
Higher-$z$ Subsample & 3.0 &           1.8 &       $55\pm5$ &                   $3.48\pm0.31$ &                    $18.3\pm1.2$ &                            $52.3\pm4.6$ &                   $2.9\pm0.3$ &                   $1.12\pm0.51$ &                     $5.9\pm2.7$ &                            $16.9\pm7.7$

%% file: half_light_radii.tex
Full Sample & 3.4\arcmin\ [3.7\arcmin] & 3.5\arcmin\ [3.8\arcmin] & 2.4\arcmin\ [2.8\arcmin] & 2.3\arcmin\ [2.8\arcmin] & 2.6\arcmin\ [3.1\arcmin] & 2.8\arcmin\ [3.3\arcmin] & 3.0\arcmin\ [3.6\arcmin] \\
Lower-z Subsample & 3.7\arcmin\ [4.0\arcmin] & 3.3\arcmin\ [3.7\arcmin] & 2.5\arcmin\ [2.9\arcmin] & 2.0\arcmin\ [2.6\arcmin] & 2.6\arcmin\ [3.0\arcmin] & 2.8\arcmin\ [3.2\arcmin] & 3.0\arcmin\ [3.5\arcmin] \\
Higher-z Subsample & 2.3\arcmin\ [2.7\arcmin] & 3.3\arcmin\ [3.7\arcmin] &  - & - & 3.3\arcmin\ [3.8\arcmin] & 3.2\arcmin\ [3.8\arcmin] & 3.2\arcmin\ [3.8\arcmin] 